\documentclass{pasj01}
\usepackage[switch,mathlines]{lineno}
\Received{$\langle$reception date$\rangle$}
\Accepted{$\langle$acception date$\rangle$}
\Published{$\langle$publication date$\rangle$}
\begin{document}

\title{3\,mm Spectroscopic Observations of Massive Star-Forming Regions with IRAM 30-m}
\author{Xuefang Xu$^{1,2}$,
Junzhi Wang$^{3}$, 
Qian Gou$^{1,2}$, 
Juan Li$^{4, 5}$,
Donghui Quan$^{6}$,
Di Li$^{7, 8}$, 
Fei Li$^{9}$, 
Chunguo Duan$^{1}$,
and
Juncheng Lei$^{1}$}
\altaffiltext{1}{School of Chemistry and Chemical Engineering, Chongqing University, 
Chongqing 401331, China}
\altaffiltext{2}{Chongqing Key Laboratory of Chemical Theory and Mechanism, Chongqing 401331, China}
\altaffiltext{3}{Guangxi Key Laboratory for Relativistic Astrophysics, Department of Physics, 
	Guangxi University, Nanning 530004, China}
\altaffiltext{4}{Shanghai Astronomical Observatory, Chinese Academy of Sciences, 
80 Nandan Road, Shanghai 200030, China}
\altaffiltext{5}{Key Laboratory of Radio Astronomy, Chinese Academy of Sciences, China}
\altaffiltext{6}{Research Center for Intelligent Computing Platforms, Zhejiang Laboratory, Hangzhou 311100, China}
\altaffiltext{7}{Department of Astronomy, Tsinghua University, Beijing 100084, China}
\altaffiltext{8}{National Astronomical Observatories, Chinese Academy of Sciences, 
	Beijing 100012, China}
\altaffiltext{9}{School of Astronomy and Space Science, Nanjing University, Nanjing 210093, China}
\email{ junzhiwang@gxu.edu.cn, qian.gou@cqu.edu.cn}


\maketitle

\begin{abstract}
Broadband spectroscopic observations with high sensitivity provide an unbiased way to 
detect emissions of molecules in space. 
We present deep observations from 105.8\,GHz to 113.6\,GHz 
toward 50 Galactic massive star-forming regions 
using IRAM 30-m millimeter telescope, with noise levels ranging from 6 to 29\,mK 
at frequency channel spacing of 195\,kHz, which corresponds to $\sim$ 0.54~km~s$^{-1}$ at 110\,GHz.
Totally, 27 molecular species have been identified, of which 16 are complex organic molecules. 
The related parameters, such as peak temperature, integrated intensity, 
and line width of the identified molecular lines were obtained. 
The line widths of the chemically related molecules show strong positive correlations, suggesting they likely originate from similar gases within star-forming regions. This work highlights the fundamental properties of the detected molecular lines and offers a valuable dataset for further studies on the astrochemical evolution of molecules in massive star-forming cores.
\end{abstract}

\section{Introduction}\label{sec:intro}

Massive star-forming regions, the cradles of high mass stars (M $>$ 8~M$_{\solar}$), 
are one kind of the most chemically rich sources in the Milky Way
~\citep{Schilke:97,Gibb:00,Belloche:13,Belloche:17,Belloche:19}. 
Information on the chemical condition of different massive star-forming regions would advance 
our understanding of the high-mass star formation~\citep{Remijan:03,Motte:18,van:21}. 
For example, Orion KL~\citep{Tercero:11,Feng:15,Bernal:21,Wilkins:22} 
and Sgr B2~\citep{Bonfand:19,Li:21,Kolesnikova:22}, 
where approximately 90\% detected interstellar molecules were first observed, 
have been found to be natural laboratories for the inventory and formation mechanism of molecules. 
Besides, other massive star-forming regions at different stages, 
such as W3(H$_2$O), W3(OH), W33A, W43, W51, W75N, ON1, Cep E, AFGL2591, 
have been conducted with molecular censuses and investigated possible formation pathways of molecules
~\citep{van:00,Fontani:07,Foster:11,Vasyunina:14,Coletta:20}. 
However, the abundances of most complex molecules in interstellar space continue to challenge current understanding. These complexities cannot be fully explained by existing chemical models, which include standard processes, such as ion-molecule and electron recombination reactions, along with surface chemistry~\citep{Ceccarelli:23}. Even the ubiquitous molecule methanol presents a challenge. 
To achieve a comprehensive understanding of interstellar chemistry, 
it is essential to test chemical models against observations. 

Broadband spectroscopic observations with high sensitivity provide 
an effective way to obtain the chemical composition of a particular region~\citep{Bonfand:19,Jorgensen:20}, 
from which chemical models can be then tested and further improved. 
Rather than targeted searches with high-priority, 
broadband observations are unbiased and self-consistent in determining 
the true compositions of interstellar clouds by allowing the acquisition of spectral data over large 
frequency ranges and sample many transitions of each molecule~\citep{Kolesnikova:22}. 
Radio astronomical observatories, 
such as IRAM 30-m (Institute de Radioastronomie Millim{\'e}trique), Yebes 40\,m, 
ALMA (Atacama Large Millimeter/submillimeter Array), 
lead to a dramatic increase in the number of publicly available broadband spectroscopic 
observations of various sources, for example, low- and high- mass star-forming regions, 
the envelope of carbon-rich star~\citep{Cernicharo:22,Margules:22,Qin:22,Pardo:22}. 
The analysis and interpretation of these spectra are important goals in the field of astrochemistry. 

In the present work, we focused on the identification of molecular species in 50 massive 
star-forming regions using IRAM's spectroscopic observations between 105.8 and 113.6\,GHz. 
In particular, among the 50 target sources, 25 sources (marked by \footnotemark[$\dag$] in Table~\ref{table:sour}) 
have not been conducted with spectroscopic observations before, 
which provides new data to gain more comprehensive view of their possible chemical and physical parameters.     
Weeds~\citep{Maret:11} was used to identify molecular species for our spectroscopic observations.
This paper is organized as follows. 
Section~\ref{sec:obda} presents the spectroscopic observations and detailed data reduction. 
The detected 30 species and radio recombination lines in our sample, 
and notes for individual sources are showed in Section~\ref{sec:res}. 
Section~\ref{sec:dis} is the discussion. The main results are summarized in Section~\ref{sec:sum}.   

\begin{table*}
\tbl{Properties of high-mass star forming regions.}{
\centering
\tiny
\begin{tabular}{llllcccc}    
\hline
\hline
Source & Alias & R.A. & Dec. & RMS & On-source & V$_{sys}$ & D \\
{} & {} & (hh:mm:ss) & (dd:mm:ss) & (mK) & time (minutes) & (km~s$^{-1}$) & (k pc) \\
\textbf{(1)} & \textbf{(2)} & \textbf{(3)} & \textbf{(4)} & \textbf{(5)} & \textbf{(6)} & \textbf{(7)} & \textbf{(8)} \\
\hline
G000.67$-$00.03 & Sgr B2                         & 17:47:20.00 & $-$28:22:40.0 & 29 & 5.7 & 62 & 0.2 \\
G005.88$-$00.39 &                                     & 18:00:30.31 & $-$24:04:04.5 & 26 & 5.7 & 9 & 5.3 \\
\footnotemark[$\dag$]G009.62$+$00.19 &        & 18:06:14.66 & $-$20:31:31.7 & 19 & 11.3 & 2 & 3.3 \\
G010.47$+$00.02 &                                    & 18:08:38.23 & $-$19:51:50.3 & 19 & 11.3 & 69 & 1.6 \\
G010.62$-$00.38 & W31                           & 18:10:28.55 & $-$19:55:48.6 & 23 & 8.5 & $-$3 & 3.8 \\
\footnotemark[$\dag$]G011.49$-$01.48 &        & 18:16:22.13 & $-$19:41:27.2 & 25 & 5.7 & 11 & 7.1 \\
\footnotemark[$\dag$]G011.91$-$00.61 &        & 18:13:58.12 & $-$18:54:20.3 & 14 & 11.3 & 37 & 5.1 \\
\footnotemark[$\dag$]G012.80$-$00.20 &        & 18:14:14.23 & $-$17:55:40.5 & 16 & 11.3 & 34 & 5.5 \\
G012.88$+$00.48 & IRAS 18089$-$1732 & 18:11:51.42 & $-$17:31:29.0 & 14 & 11.3 & 31 & 5.9 \\
\footnotemark[$\dag$]G012.90$-$00.24 &                                    & 18:14:34.42 & $-$17:51:51.9 & 25 & 5.7 & 36 & 5.9 \\
\footnotemark[$\dag$]G012.90$-$00.26 &                                    & 18:14:39.57 & $-$17:52:00.4 & 18 & 8.5 & 39 & 5.9 \\
\footnotemark[$\dag$]G014.33$-$00.64 &        & 18:18:54.67 & $-$16:47:50.3 & 27 & 5.7 & 22 & 7.2 \\
G015.03$-$00.67 & M17                          & 18:20:24.81 & $-$16:11:35.3 & 19 & 5.7 & 22 & 6.4 \\
\footnotemark[$\dag$]G016.58$-$00.05 &       & 18:21:09.08 & $-$14:31:48.8 & 16 & 8.5 & 60 & 5.0 \\
\footnotemark[$\dag$]G023.00$-$00.41 &        & 18:34:40.20 & $-$09:00:37.0 & 14 & 11.3 & 80 & 4.5 \\
\footnotemark[$\dag$]G023.44$-$00.18 &        & 18:34:39.19 & $-$08:31:25.4 & 14 & 11.3 & 97 & 3.7 \\
\footnotemark[$\dag$]G027.36$-$00.16 &        & 18:41:51.06 & $-$05:01:43.4 & 13 & 11.3 & 92 & 3.9 \\
\footnotemark[$\dag$]G028.86$+$00.06 &       & 18:43:46.22 & $-$03:35:29.6 & 12 & 11.3 & 100 & 4.0 \\
G029.95$-$00.01 & W43S                        & 18:46:03.74 & $-$02:39:22.3 & 18 & 11.3 & 98 & 4.6 \\
\footnotemark[$\dag$]G031.28$+$00.06 &       & 18:48:12.39 & $-$01:26:30.7 & 17 & 11.3 & 109 & 5.2 \\
G031.58$+$00.07 & W43Main                 & 18:48:41.68 & $-$01:09:59.0 & 15 & 11.3 & 96 & 4.9 \\
\footnotemark[$\dag$]G032.04$+$00.05 &       & 18:49:36.58 & $-$00:45:46.9 & 10 & 19.8 & 97 & 4.8 \\
\footnotemark[$\dag$]G034.39$+$00.22 &       & 18:53:18.77 & $+$01:24:08.8 & 8 & 39.0 & 57 & 7.1 \\
\footnotemark[$\dag$]G035.02$+$00.34 &       & 18:54:00.67 & $+$02:01:19.2 & 8 & 39.0 & 52 & 6.5 \\
\footnotemark[$\dag$]G035.19$-$00.74  &       & 18:58:13.05 & $+$01:40:35.7 & 13 & 14.1 & 30 & 6.6 \\
\footnotemark[$\dag$]G035.20$-$01.73  &       & 19:01:45.54 & $+$01:13:32.5 & 10 & 22.6 & 42 & 5.9 \\
\footnotemark[$\dag$]G037.43$+$01.51 &       & 18:54:14.35 & $+$04:41:41.7 & 9 & 34.5 & 41 & 6.9 \\
G043.16$+$00.01 & W49N                       & 19:10:13.41 & $+$09:06:12.8 & 14 & 19.8 & 10 & 7.6 \\
G043.79$-$00.12 & OH 43.8$-$0.1           & 19:11:53.99 & $+$09:35:50.3 & 10 & 25.5 & 44 & 5.7 \\
G049.48$-$00.36 & W51 IRS2                 & 19:23:39.82 & $+$14:31:05.0 & 15 & 11.3 & 56 & 6.3 \\
G049.48$-$00.38 & W51M                       & 19:23:43.87 & $+$14:30:29.5 & 12 & 11.3 & 58 & 6.3 \\
\footnotemark[$\dag$]G059.78$+$00.06 &       & 19:43:11.25 & $+$23:44:03.3 & 11 & 14.1 & 25 & 7.5 \\
G069.54$-$00.97 & ON1                          & 20:10:09.07 & $+$31:31:36.0 & 10 & 22.6 & 12 & 7.8 \\
\footnotemark[$\dag$]G075.76$+$00.33 &       & 20:21:41.09 & $+$37:25:29.3 & 8 & 22.6 & $-$9 & 8.2 \\
G078.12$+$03.63 & IRAS 20126+4104    & 20:14:26.07 & $+$41:13:32.7 & 16 & 11.3 & $-$4 & 8.1 \\
G081.75$+$00.59 & DR21                        & 20:39:01.99 & $+$42:24:59.3 & 9 & 22.6 & $-$3 & 8.2 \\
G081.87$+$00.78 & W75N                       & 20:38:36.43 & $+$42:37:34.8 & 11 & 19.8 & 7 & 8.2 \\
\footnotemark[$\dag$]G092.67$+$03.07 &       & 21:09:21.73 & $+$52:22:37.1 & 12 & 14.1 & $-$5 & 8.5 \\
G109.87$+$02.11 & Cep A                        & 22:56:18.10 & $+$62:01:49.5 & 10 & 22.6 & $-$7 & 8.6 \\
G111.54$+$00.77 & NGC 7538                 & 23:13:45.36 & $+$61:28:10.6 & 6 & 85.5 & $-$57 & 9.6  \\
G121.29$+$00.65 & L1287                       & 00:36:47.35 & $+$63:29:02.2 & 6 & 60.0 & $-$23 & 8.8 \\
G123.06$-$06.30 & NGC281                    & 00:52:24.70 & $+$56:33:50.5 & 8 & 64.5 & $-$30 & 10.1 \\
G133.94$+$01.06 & W3OH                      & 02:27:03.82 & $+$61:52:25.2 & 13 & 11.3 & $-$47 & 9.8 \\
G168.06$+$00.82 & IRAS 05137+3919    & 05:17:13.74 & $+$39:22:19.9 & 17 & 14.1 & $-$27 & 15.9 \\
\footnotemark[$\dag$]G176.51$+$00.20 &       & 05:37:52.14 & $+$32:00:03.9 & 6 & 117.0 & $-$17 & 9.3 \\
\footnotemark[$\dag$]G183.72$-$03.66 &        & 05:40:24.23 & $+$23:50:54.7 & 5 & 87.0 & 3 & 10.0 \\
G188.94$+$00.88 & S 252                         & 06:08:53.35 & $+$21:38:28.7 & 9 & 60.0 & 8 & 10.4 \\
G192.60$-$00.04 & S 255                          & 06:12:54.02 &$+$17:59:23.3 & 9 & 57.0 & 6 & 9.9 \\
G209.00$-$19.38 & Orion Nebula             & 05:35:15.80 & $-$05:23:14.1 & 6 & 84.0 & 3 & 8.6 \\
\footnotemark[$\dag$]G232.62$+$00.99 &       & 07:32:09.78 & $-$16:58:12.8 & 8 & 75.0 & 21 & 9.4 \\ 
\hline
\end{tabular}}\label{table:sour}
\begin{tabnote}
{\bf Note.} \footnotemark[$\dag$]  Sources have not been conducted with spectroscopic observations before. 
Columns (1) and (2) give the Galactic source name/coordinates and an alias, when appropriate. 
Right Ascension and Declination (J2000) are listed in columns (3) and (4). 
Columns (5), (6), (7), and (8) list RMS, on-source time, systemic velocities and distances of sources from the galactic centre, respectively. 
Values in columns (7) and (8) are taken from~\citet{Reid:14} and~\citet{Li:22}, respectively. 
\end{tabnote}
\end{table*}

\begin{tiny}
\setlength{\tabcolsep}{2.0pt}
\begin{longtable}{clccc|clccc}   
	\caption{Rotational temperatures and column densities of detected molecules using in Weeds} \label{table:paraweeds}\\    
	\hline
	\hline
	Source & Species& Rotational & Column & Velocity & Source & Species& Rotational & Column & Velocity \\ 
	{} & {}  & Temperature  & Density  & offset  & 
	{} & {}  & Temperature  & Density  & offset  \\
	{} & {}  & (K) & (cm$^{-2}$) & (km~s$^{-1}$) &   
	{} & {}  & (K) & (cm$^{-2}$) & (km~s$^{-1}$) \\
	\textbf{(1)} & \textbf{(2)} & \textbf{(3)} & \textbf{(4)} & \textbf{(5)} & \textbf{(1)} & \textbf{(2)} & \textbf{(3)} & \textbf{(4)} & \textbf{(5)} \\
	\hline
	\endfirsthead
	\hline
	\hline
	Source & Species& Rotational & Column & Velocity & Source & Species& Rotational & Column & Velocity \\ 
	{} & {}  & Temperature  & Density  & offset  & 
	{} & {}  & Temperature  & Density  & offset  \\
	{} & {}  & (K) & (cm$^{-2}$) & (km~s$^{-1}$) &   
	{} & {}  & (K) & (cm$^{-2}$) & (km~s$^{-1}$) \\
	\textbf{(1)} & \textbf{(2)} & \textbf{(3)} & \textbf{(4)} & \textbf{(5)} & \textbf{(1)} & \textbf{(2)} & \textbf{(3)} & \textbf{(4)} & \textbf{(5)} \\
	\hline
	\endhead
	\endfoot
	\hline
	\multicolumn{8}{l}{{\bf Note.} This is only a subset of the whole table, which is available in the supplementary material.}
	\endlastfoot
	{} & NH$_{2}$CHO & 45 & 1.8$\times$10$^{15}$ & 0.2 & {} & CCS & 45 & 2.0$\times$10$^{13}$ & 1.0 \\
	{} & CCS & 45 & 2.5$\times$10$^{14}$ & 0.5 & {} & HC$_{5}$N & 45 & 1.5$\times$10$^{13}$ & 1.0 \\
	{} & C$_{2}$H$_{5}$CN & 45 & 2.0$\times$10$^{14}$ & 0. & {} & $^{34}$SO & 45 & 3.7$\times$10$^{13}$ & -0.5 \\
	{} & HC$_{5}$N & 45 & 8.5$\times$10$^{13}$ & 0. & {} & CH$_{3}$OH & 85 & 2.2$\times$10$^{15}$ & 0. \\ 
	{} & C$_{2}$H$_{3}$CN & 45 & 8.0$\times$10$^{14}$ & 0.5 & {} & t-HCOOH & 85 & 1.2$\times$10$^{14}$ & 1.0 \\ 
	{} & C$_{2}$H$_{5}$OH & 45 & 3.5$\times$10$^{15}$ & 0. & {} & $^{13}$CN & 85 & 1.0$\times$10$^{14}$ & 1.0 \\  
	{} & CH$_{3}$OCH$_{3}$ & 45 & 9.0$\times$10$^{15}$ & 0. & {} & HC$_{3}$N & 85 & 3.0$\times$10$^{14}$ & 1.0 \\ 
	{} & OC$^{34}$S & 45 & 1.5$\times$10$^{15}$ & 0. & G035.02$+$00.34 & SO & 45 & 1.5$\times$10$^{15}$ & 0.5 \\
	{} & CH$_{3}$OH & 45 & 3.3$\times$10$^{16}$ & -2.0 & {} & OCS & 45 & 4.7$\times$10$^{14}$ & 0.5 \\ 
	{} & CH$_{3}$C$^{15}$N & 45 & 8.5$\times$10$^{13}$ & -2.0 & {} & C$^{18}$O & 85 & 2.2$\times$10$^{17}$ & 0. \\  
	{} & $^{13}$CH$_{3}$CN & 45 & 1.0$\times$10$^{14}$ & 0. & {} & HNCO & 45 & 7.5$\times$10$^{13}$ & 1.0 \\   
	{} & CH$_{3}$SH & 45 & 4.5$\times$10$^{15}$ & 0. & {} & $^{13}$CO & 85 & 1.7$\times$10$^{18}$ & 0. \\ 
	{} & CH$_{3}$OCHO & 45 & 1.0$\times$10$^{16}$ & 0. & {} & CH$_{3}$CN & 45 & 3.8$\times$10$^{13}$ & 0. \\    
	G000.67$-$00.03 & SO$_{2}$ & 45 & 1.6$\times$10$^{16}$ & -2.0 & {} & CH$_{3}$CHO & 45 & 1.7$\times$10$^{14}$ & 0. \\  
	{} & t-HCOOH & 45 & 1.5$\times$10$^{15}$ & 0. & {} & C$^{17}$O & 45 & 3.3$\times$10$^{16}$ & 0.2 \\  
	{} & SiS & 45 & 9.0$\times$10$^{14}$ & 0. & {} & CN & 45 & 3.5$\times$10$^{15}$ & 1.0 \\ \cline{6-10} 
	{} & HC$_{3}$N & 45 & 4.0$\times$10$^{15}$ & 0. & {} & NH$_{2}$CHO & 45 & 2.3$\times$10$^{13}$ & 3.0 \\  
	{} & SO & 45 & 1.2$\times$10$^{16}$ & 0. & {} & CCS & 45 & 2.7$\times$10$^{13}$ & 4.0 \\    
	{} & HC$_{3}$N, v$_{7}$=1 & 45 & 1.0$\times$10$^{17}$ & 0. & {} & HC$_{5}$N & 45 & 2.2$\times$10$^{13}$ & 4.5 \\ 
	{} & OCS & 45 & 1.3$\times$10$^{16}$ & 0. & {} & $^{34}$SO & 45 & 4.5$\times$10$^{13}$ & 2.0 \\  
	{} & HNCO & 45 & 1.2$\times$10$^{15}$ & 0. & {} & CH$_{3}$OCH$_{3}$ & 45 & 1.5$\times$10$^{15}$ & 2.0 \\ 
	{} & C$^{18}$O & 45 & 5.1$\times$10$^{17}$ & 3.0 & {} & OC$^{34}$S & 45 & 1.0$\times$10$^{14}$ & 2.0 \\ 
	{} & $^{13}$CO & 85 & 8.0$\times$10$^{18}$ & 3.0 & {} & CH$_{3}$OH & 85 & 2.7$\times$10$^{16}$ & 1.5 \\                   
	{} & CH$_{3}$$^{13}$CN & 45 & 1.2$\times$10$^{14}$ & 0. & {} & CH$_{3}$OCHO & 85 & 1.2$\times$10$^{15}$ & 3.0 \\                 
	{} & CH$_{3}$CN & 45 & 1.5$\times$10$^{15}$ & 0. & {} & SO$_{2}$ & 85 & 1.7$\times$10$^{15}$  & 1.0 \\                     
	{} & CH$_{3}$CHO & 45 & 2.6$\times$10$^{15}$ & 0. & {} & t-HCOOH & 85 & 3.0$\times$10$^{14}$ & 4.0 \\                      
	{} & C$^{17}$O & 45 & 1.6$\times$10$^{17}$ & 3.5 & G035.19$-$00.74 & $^{13}$CN & 85 & 2.0$\times$10$^{14}$ & 4.0 \\       
	{} & CN & 85 & 1.0$\times$10$^{15}$ & -2.0 & {} & HC$_{3}$N & 85 & 6.0$\times$10$^{14}$ & 4.0 \\ \cline{1-5}                    
	{} & CCS & 45 & 5.2$\times$10$^{13}$ & 0. & {} & SO & 45 & 1.8$\times$10$^{15}$ & 4.0 \\                           
	{} & HC$_{5}$N & 45 & 9.5$\times$10$^{13}$ & -0.5 & {} & OCS & 45 & 1.3$\times$10$^{15}$ & 4.0 \\                          
	{} & C$_{2}$H$_{3}$CN & 45 & 1.1$\times$10$^{14}$ & -0.5 & {} & C$^{18}$O & 85 & 2.0$\times$10$^{17}$ & 4.0 \\                    
	{} & CH$_{3}$OH & 45 & 6.5$\times$10$^{15}$ & 0.5 & {} & HNCO & 45 & 2.5$\times$10$^{14}$ & 4.0 \\                         
	{} & SO$_{2}$ & 45 & 1.0$\times$10$^{18}$ & 0. & {} & C$^{15}$N & 45 & 3.0$\times$10$^{13}$ & 4.0 \\                    
	{} & CH$_{3}$C$_{3}$N & 45 & 7.0$\times$10$^{13}$ & 0. & {} & $^{13}$CO & 85 & 3.0$\times$10$^{18}$ & 4.0 \\                    
	{} & $^{13}$CN & 45 & 3.2$\times$10$^{14}$ & -0.7 & {} & CH$_{3}$CN & 45 & 1.7$\times$10$^{14}$ & 3.5 \\                   
	{} & HC$_{3}$N & 85 & 8.5$\times$10$^{15}$ & 0. & {} & CH$_{3}$CHO & 45 & 5.2$\times$10$^{14}$ & 4.0 \\                  
	{} & SO & 45 & 1.0$\times$10$^{16}$ & 0. & {} & C$^{17}$O & 45 & 2.3$\times$10$^{16}$ & 4.0 \\                    
	G005.88$-$00.39 & HC$_{3}$N, v$_{7}$=1 & 85 & 1.5$\times$10$^{15}$ & 0.5 & {} & CN & 45 & 3.0$\times$10$^{15}$ & 4.0 \\ \cline{6-10}                          
	{} & OCS & 45 & 3.0$\times$10$^{15}$ & 0.5 & {} & CCS & 45 & 8.5$\times$10$^{12}$ & 2.0 \\                          
	{} & C$^{18}$O & 85 & 4.3$\times$10$^{17}$ & 0. & {} & CH$_{3}$OH & 85 & 1.0$\times$10$^{15}$ & 1.5 \\                   
	{} & HNCO & 45 & 2.5$\times$10$^{14}$ & 0. & {} & t-HCOOH & 85 & 1.0$\times$10$^{14}$ & 2.0 \\                      
	{} & C$^{15}$N & 45 & 8.5$\times$10$^{13}$ & 0. & {} & $^{13}$CN & 85 & 1.2$\times$10$^{14}$ & 2.0 \\                    
	{} & $^{13}$CO & 85 & 9.2$\times$10$^{18}$ & 0. & {} & HC$_{3}$N & 85 & 1.5$\times$10$^{14}$ & 2.0 \\                    
	{} & CH$_{3}$CN & 45 & 7.5$\times$10$^{14}$ & 0. & {} & SO & 45 & 1.0$\times$10$^{15}$ & 2.0 \\                           
	{} & CH$_{3}$OCHO & 45 & 3.4$\times$10$^{14}$ & -0.5 & {} & OCS & 45 & 4.0$\times$10$^{14}$ & 2.0 \\                          
	{} & CH$_{3}$OCH$_{3}$ & 45 & 1.0$\times$10$^{15}$ & 1.0 & G035.20$-$01.73 & C$^{18}$O & 85 & 1.1$\times$10$^{17}$ & 2.0 \\       
	{} & C$^{17}$O & 45 & 5.5$\times$10$^{16}$ & 0. & {} & HNCO & 45 & 4.5$\times$10$^{13}$ & 2.0 \\                         
	{} & CN & 85 & 3.0$\times$10$^{16}$ & 0. & {} & C$^{15}$N & 45 & 4.0$\times$10$^{13}$ & 2.0 \\ \cline {1-5}                    
	{} & NH$_{2}$CHO & 45 & 7.0$\times$10$^{13}$ & 1.0 & {} & $^{13}$CO & 85 & 1.7$\times$10$^{18}$ & 2.0 \\                    
	{} & CCS & 45 & 2.5$\times$10$^{13}$ & 1.5 & {} & CH$_{3}$CN & 45 & 4.1$\times$10$^{13}$ & 2.0 \\        
	{} & HC$_{5}$N & 45 & 2.0$\times$10$^{13}$ & 1.5 & {} & CH$_{3}$CHO & 45 & 1.0$\times$10$^{14}$ & 2.0 \\                  
	{} & C$_{2}$H$_{5}$OH & 45 & 4.1$\times$10$^{14}$ & 2.0 & {} & C$^{17}$O & 45 & 2.0$\times$10$^{16}$ & 2.0 \\                    
	{} & $^{34}$SO & 45 & 1.1$\times$10$^{14}$ & 3.0 & {} & CN & 45 & 3.0$\times$10$^{15}$ & 1.0 \\ \cline {6-10} 
	{} & CH$_{3}$OCH$_{3}$ & 45 & 1.0$\times$10$^{15}$ & 1.0 & {} & CCS & 45 & 1.7$\times$10$^{13}$ & 3.0 \\                          
	{} & OC$^{34}$S & 45 & 2.3$\times$10$^{14}$ & 2.0 & {} & HC$_{5}$N & 45 & 1.2$\times$10$^{13}$ & 3.5 \\                    
	{} & CH$_{3}$OH & 85 & 1.2$\times$10$^{16}$ & 2.5 & {} & $^{34}$SO & 45 & 3.3$\times$10$^{13}$ & 2.0 \\                    
	{} & C$_{2}$H$_{5}$CN & 45 & 7.3$\times$10$^{13}$ & 3.0 & {} & CH$_{3}$OH & 85 & 4.0$\times$10$^{15}$ & 2.5 \\                   
	{} & $^{13}$CH$_{3}$CN & 45 & 5.5$\times$10$^{12}$ & 2.0 & {} & CH$_{3}$OCHO & 45 & 4.7$\times$10$^{14}$ & 3.5 \\                 
	{} & CH$_{3}$OCHO & 45 & 1.3$\times$10$^{15}$ & 2.0 & {} & SO$_{2}$ & 85 & 4.5$\times$10$^{14}$ & 3.0 \\                     
	{} & t-HCOOH & 45 & 6.7$\times$10$^{13}$ & 1.5 & {} & t-HCOOH & 85 & 1.5$\times$10$^{14}$ & 3.0 \\                      
	G009.62$+$00.19 & $^{13}$CN & 45 & 1.2$\times$10$^{14}$ & 2.3 & {} & $^{13}$CN & 85 & 1.0$\times$10$^{14}$ & 2.5 \\                    
	{} & HC$_{3}$N & 85 & 9.2$\times$10$^{14}$ & 2.0 & {} & HC$_{3}$N & 85 & 1.7$\times$10$^{14}$ & 3.0 \\                    
	{} & SO & 45 & 2.0$\times$10$^{15}$ & 2.5 & G037.43$+$01.51 & SO & 45 & 1.3$\times$10$^{15}$ & 3.0 \\              
	{} & HC$_{3}$N, v$_{7}$=1 & 85 & 6.5$\times$10$^{14}$ & 1.0 & {} & OCS & 45 & 4.7$\times$10$^{14}$ & 3.0 \\                          
	{} & OCS & 45 & 2.7$\times$10$^{15}$ & 2.5 & {} & HNCO & 45 & 7.0$\times$10$^{13}$ & 2.5 \\                         
	{} & HNCO & 45 & 1.2$\times$10$^{14}$ & 2.0 & {} & C$^{18}$O & 85 & 1.9$\times$10$^{17}$ & 3.0 \\                    
	{} & C$^{18}$O & 85 & 4.3$\times$10$^{17}$ & 2.0 & {} & C$^{15}$N & 45 & 1.0$\times$10$^{13}$ & 2.5 \\                    
	{} & C$^{15}$N & 45 & 1.5$\times$10$^{12}$ & 2.5 & {} & $^{13}$CO & 85 & 4.4$\times$10$^{18}$ & 3.0 \\                    
	{} & $^{13}$CO & 85 & 8.5$\times$10$^{18}$ & 2.5 & {} & CH$_{3}$CN & 45 & 5.0$\times$10$^{13}$ & 3.0 \\                   
	{} & CH$_{3}$CN & 45 & 2.0$\times$10$^{14}$ & 2.0 & {} & CH$_{3}$OCH$_{3}$ & 45 & 1.0$\times$10$^{15}$ & 3.0 \\            
	{} & C$_{2}$H$_{3}$CN & 45 & 1.1$\times$10$^{13}$ & 2.5 & {} & CH$_{3}$CHO & 45 & 1.3$\times$10$^{14}$ & 3.0 \\                  
	{} & CH$_{3}$CHO & 45 & 1.4$\times$10$^{14}$ & 2.0 & {} & C$^{17}$O & 45 & 2.6$\times$10$^{16}$ & 3.0 \\                    
	{} & C$^{17}$O & 45 & 7.0$\times$10$^{13}$ & 1.5 & {} & CN & 85 & 1.0$\times$10$^{15}$ & 3.0 \\ 
	{} & CN & 85 & 1.0$\times$10$^{16}$ & 2.2 & {} & {} & {} & {} & {} \\ \hline                          
\end{longtable}
\end{tiny}


\section{Observations and data reduction} \label{sec:obda}
\subsection{Observations} \label{sec:obser}
Broadband spectroscopic observations of 50 massive star-forming regions selected from~\citet{Reid:14} were conducted with 
the IRAM 30-m millimeter telescope at Pico Veleta, Spain, on June and October  2016, and August 2017. 
The criteria to select the target sources were detailed in~\citet{Li:22}. 
The data used in this paper was taken by the 3\,mm (E0) band of the Eight Mixer Receiver (EMIR). 
The Fourier Transform Spectrometers (FTS) cover 8\,GHz bandwidth 
and provide a spectral resolution of 195\,kHz (0.54~km~s$^{-1}$) with dual polarization.
Standard position switching mode with azimuth off of -600$''$ was adopted. 
Due to the emission at off position, 
the contamination can cause under-estimation for some strong lines with low critical density, 
which needs to be taken care only in sources like Sgr B2. 
Even for Sgr B2, it will be not important for most of the weak lines with high critical densities~\citep{Jones:12}.     
Detailed descriptions of the observations were presented in~\citet{Li:22}. 
Table~\ref{table:sour} lists the information of the 50 massive star-forming regions.   

\subsection{Data reduction} \label{sec:data}
\begin{figure*}
 \centering
  \includegraphics[width=0.65\textwidth]{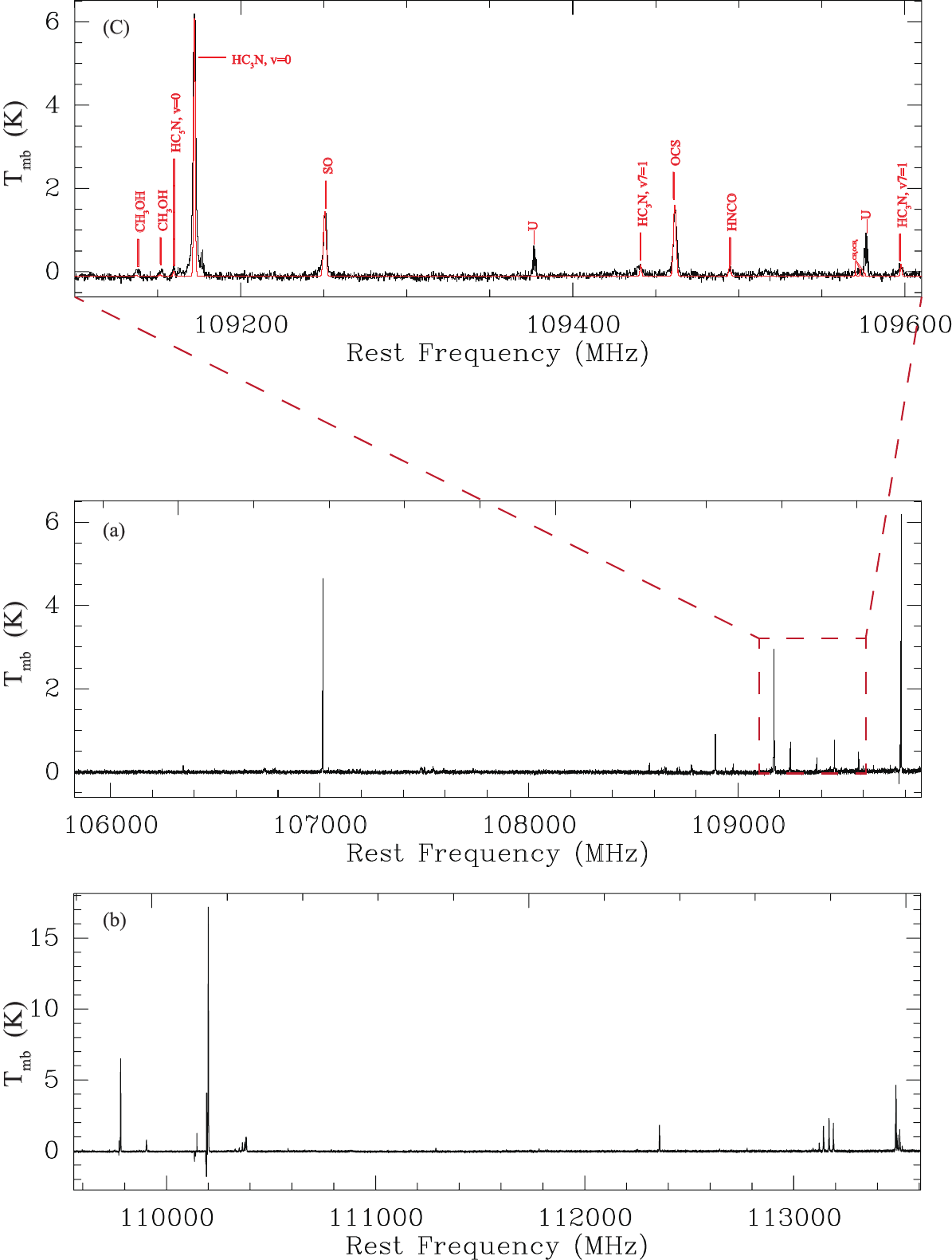}
   \caption{(a) and (b): The observed lines that have subtracted baselines in G009.62$+$00.19. 
(c) Zoomed-in view of the modeling synthetic spectrum of the observed lines. 
The red solid line are the modeling synthetic spectrum.}
\label{fig:line}           
\end{figure*}

The data reduction was conducted with the 
GILDAS (Grenoble Image and Line Data Analysis Software) software package\footnote{http://www.iram.fr/IRAMFR/GILDAS}, 
including `CLASS' (Continuum and Line Analysis Single-dish Software), 
`GREG'(Grenoble Graphic), and `Weeds'{Maret:11}.
Main steps of data reduction are described as follows. 
We first averaged the scans of spectra for the same source after checking data quality of every scan.  
Less than 0.5\% data were discarded due to poor baseline or bad channels.  
The first-order baseline was then applied on  the averaged spectra for  most of sources.
Seven sources (G009.62$+$00.19, G010.47$+$00.02, G035.20$-$01.73, G059.78$+$00.06, 
G069.54$-$00.97, G081.75$+$00.59, and G081.87$+$00.78) 
were applied second-order baseline for the appearance of the significant break. 
Since G043.16$+$00.01 and G049.48$-$00.36 show regular small standing waves, 
they were subtracted sine baselines. 
The full scan of G009.62$+$00.19 after subtracting second 
order baselines is shown in Figure~\ref{fig:line} (a) and (b). 
The observed lines in each source were then identified using Weeds. 
The line identification relied heavily on the accuracy of frequencies and line strengths 
derived from the analysis of laboratory measurements. 
Weeds was also used to model the synthetic spectra of the observed lines 
with the assumption of local thermodynamic equilibrium (LTE).
The JPL (Jet Propulsion Laboratory)~\citep{Pickett:98} 
and CDMS (Cologne Database for Molecular Spectroscopy)~\citep{Muller:05, Endres:16} 
molecular databases were used for the whole identification and spectral modeling process. 
Spectral modeling is supported by five parameters: 
source size, rotational temperature (Col. (3) in Table~\ref{table:paraweeds}), 
column density (Col. (4) in Table~\ref{table:paraweeds}), 
velocity offset (Col. (5) in Table~\ref{table:paraweeds}) with respect to 
the systemic velocity (Col. (7) in Table~\ref{table:sour}), and line width. To standardize our analysis across different distances, a uniform source size of 10$^{''}$ was applied to all observed sources. This assumption is based on typical extents of emission in similar studies~\citep{Li:22}, which suggests a reasonable compromise between resolution limits (the angular resolution, $\sim$ 24$^{''}$, of the IRAM data) and actual source sizes. This approach introduces an estimated uncertainty of about 30\% in our column density calculations. 
However, the provided parameters, including  line intensities and widths in each source, 
do not significantly change with different assumed source sizes. 
Figure~\ref{fig:line} (c) is an example for the modeling synthetic spectrum (red solid line) 
overlaid with the observed lines in G009.62$+$00.19.
Finally, Gaussian fits were performed for the identified lines to get peak intensity, 
centroid velocity, full width at half maximum (FWHM), and integrated intensity. 
The results derived with Gaussian fitting are reported in Section~\ref{sec:mld}.   

\section{Results} \label{sec:res}
\subsection{Detections of molecular species} \label{sec:mld}
Molecules identified in each source are summarized in the second, third and fourth columns of Table~\ref{table:mole}.  
Excluding isotopologues, 27 compounds have been identified in the 50 target massive star-forming regions, 
of which 16 species contain six or more atoms. 
The identified species cover 10~oxygen-bearing molecules,~5~sulfur-bearing molecules,
10~nitrogen-bearing molecules,~1~silicon-bearing molecule, and~1~ion. 
Particularly, SO, CN, HC$_{3}$N, and CH$_{3}$OH are detected in all sources, 
while CH$_{3}$CN is detected in all sources except for G168.06$+$00.82. 
On the other hand, some molecules were detected in most of sources, such as OCS (48 sources), 
HNCO (43 sources), CH$_{3}$CHO (42 sources), 
while some molecules were detected only in few sources, 
such as CH$_{3}$SH in G010.62$-$00.38, G012.80$-$00.20 and G049.48$-$00.36, 
CH$_{3}$COOH in G010.47$+$00.02, G049.48$-$00.36, and G049.48$-$00.38.  
Molecules chemically related with each other are discussed in Section~\ref{sec:rmc}. 
The only detected ion, HOCO$^{+}$, is detailed in Section~\ref{sec:hoco}.    

Lines of the detected molecules, above 3$\sigma$ level in each source, 
are listed in Table~\ref{table:moled}, 
which include species, transitions, rest frequencies, upper level energy (E$_{u}$), 
line strength (i.e. dipole-weighted transition dipole matrix elements, $\mu$$^{2}$S), 
peak intensity (T$_{mb}$), centroid velocity (V$_{\rm LSR}$), FWHM ($\Delta$V), 
and integrated intensity ($\int$T$_{mb}$dv).  
Lines in different vibrational states, and with high E$_{u}$ were also identified. 
For example,  lines of HC$_{3}$N in v$_{4}$=1, v$_{6}$=1, v$_{7}$=1, v$_{7}$=2, v$_{5}$=1/v$_{7}$=3, 
and v$_{6}$=v$_{7}$=1, as well as CH$_{3}$CN in v$_{8}$=1, and C$_{2}$H$_{5}$CN with v$_{12}$=1-A, v$_{20}$=1-A 
vibrational states with high E$_{u}$ have been observed. 

\subsection{Radio recombination lines} \label{sec:rrl}
Six radio recombination lines (RRLs) of hydrogen and  three RRLs of helium were detected in this work, 
which are listed in the column (5) of Table~\ref{table:mole}. 
H (39) $\alpha$ is the strongest among the identified RRLs, detected in 20 sources. 
The He (39) $\alpha$ was detected in 7 sources, while He (48) $\beta$ and 
He (55) $\gamma$ were separately detected in G012.80$-$00.20 and G209.00$-$19.38. 
Line parameters for the detections of H (39) $\alpha$ and He (39) $\alpha$ 
in the different sources are listed in Table~\ref{table:RRLs}. 

\subsection{Notes for individual sources} \label{sec:is}
The detected molecules in G000.67$-$00.03 (Sgr B2), G010.47$+$00.02, G049.48$-$00.36, 
and G049.48$-$00.38 cover more than 98\% of the identified species. 
The four sources are rich in 18, 25, 21, and 23 molecular species, respectively. 
On the other hand, some molecules were only detected in one or two of the four sources. 
For example, HCOCN and CH$_{3}$OCH$_{2}$OH were only detected in G010.47$+$00.02,   
while aGg'$-$(CH$_{2}$OH)$_{2}$ was detected in G010.47$+$00.02 and G049.48$-$00.38. 
The RRLs in G012.80$-$00.20 and G015.03$-$00.67 include all kinds of RRLs observed in this work. 

\subsection{Unidentified lines}           
The observed frequencies for the unidentified lines are listed in Table~\ref{table:rul}. 
The JPL and CDMS databases cannot provide any plausible carrier for these unidentified features, 
the line width of which ranges from 1.1 to 19.0\,km/s. 
These unidentified lines with different widths for the same source are credible. 
The widths of unidentified lines are within the width range of those detected lines. 
For example, line widths of the detected lines and unidentified lines in G009.62$+$00.19 
are from 0.5 to 19.1\,km/s, and from 1.5 to 5.2\,km/s, respectively.
Only three of our target sources, G005.88$-$00.39, G031.28$+$00.06, and G111.54$+$00.77, 
have no unidentified features above 3$\sigma$ level. 
In 33 sources, there are four or less than four lines above 3$\sigma$ level left for identification. 
G049.48$-$00.38 (W 51M) has the largest number of unidentified features (40 in total), 
which deserves more attention for search of new interstellar molecules. 
Table~\ref{table:rul} can be taken as a reference for follow-up in future works. 
Among these unidentified lines, H$_{2}$CCN and DCOOH might be the carriers, 
of which only one transition matches with the laboratory spectroscopic data 
in G005.88$-$00.39 and G009.62$+$00.19, respectively. 


\section{Discussion} \label{sec:dis} 
\subsection{Column Densities and Line Widths of the Detected Lines}
\begin{figure*}
	\centering
	\includegraphics[width=0.85\textwidth]{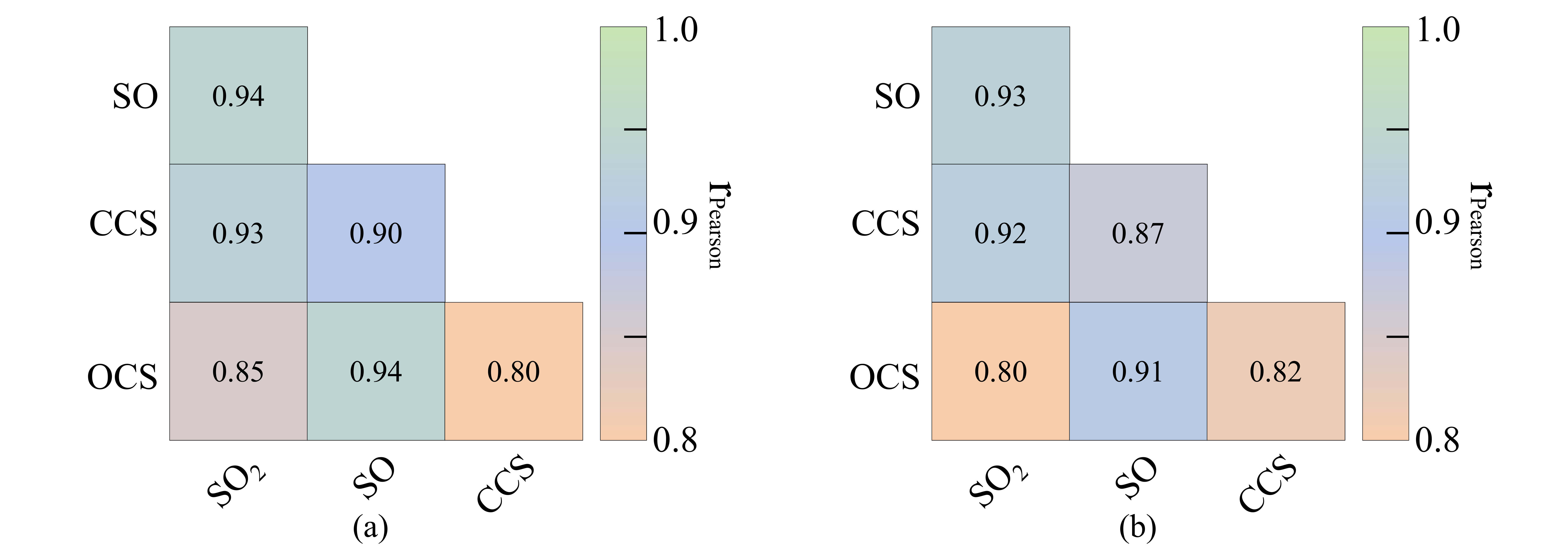}
	\caption{Correlation matrix displaying the Pearson coefficients for (a) line widths and (b) column densities of the detected SO, CCS, OCS, and SO$_{2}$ in our sample.}
	\label{fig:wid}           
\end{figure*}

To further elucidate the underlying similarities and correlations among various molecules, the Pearson's coefficient ($r_{Pearson}$) was employed to  separately compare column densities (column (4) in Table~\ref{table:paraweeds}) and line widths (FWHMs, column (8) in Table~\ref{table:moled}) of the detected lines, as derived from the line-fitting procedure detailed in Section~\ref{sec:data}. Figure~\ref{fig:wid} visualizes this comparative analysis of line widths and column densities. The strong positive correlations ($r_{Pearson}{}>{}0.80$) observed in both column densities and line widths provide further validation that SO, CCS, OCS, and SO$_{2}$, are chemically linked with each other, as discussed in Section~\ref{sec:so}. Using a similar approach,~\citet{Fontani:23} reported poor correlations among the FWHMs of CCS, SO$_{2}$, and OCS across different evolutionary stages of 15 high-mass star-forming cores. These correlations could potentially become more robust if analyzed with identical data points align with that of this work. Furthermore, the strong correlations identified in this work suggest that these molecules, along with other chemically related species outlined in Sections~\ref{sec:ch3oh} $-$~\ref{sec:cn}, likely originate from similar gaseous environments within star-forming regions~\citep{Coletta:20}.

\subsection{Related Molecules in Chemistry} \label{sec:rmc}
One way to investigate possible formation pathways of molecules is through 
statistically analyzing abundances of related molecules~\citep{Pilling:11,Peng:22}. 
The related molecules, particularly complex organic molecules (COMs), 
are of considerable interest for studying the astro-chemical evolution where they reside~\citep{Herbst:09}, 
because COMs have the potential to form prebiotic molecules such as amino acids, 
sugars, and nucleobases~\citep{Gorai:20}. 
Following the convention of ~\citet{Herbst:09}, 
a COM is defined as a molecule containing six or more atoms 
and 16 species of COMs have been detected in this work.  
The detection rates of the related molecules are listed in Table~\ref{table:dr}, which is convenient to study the potential formation routes of molecules and the physical properties, such as temperature and mass, of the environments where they reside.

\subsubsection{SO, CCS, OCS, and SO$_{2}$} \label{sec:so} 
SO, CCS, OCS, and SO$_{2}$ are usually used to trace dense gas~\citep{Benson:98,Holdship:19,Xie:21}, 
and may be chemically correlated with each other~\citep{Herpin:09, Li:15,Vidal:18}. 
For example,~\citet{Vastel:18} focused on 21 sulfur-bearing species, including SO, CCS, OCS, and SO$_{2}$, 
to investigate sulfur chemistry in the L1544 prestellar core. 
Additionally,~\citet{Fontani:23} modeled the chemistry of sulfur during the evolution of high-mass star-forming cores 
according to the detected lines of SO, SO$^{+}$, NS, C$^{34}$S, $^{13}$CS, SO$_{2}$, CCS, H$_{2}$S, HCS$^{+}$, OCS, H$_{2}$CS, and CCCS. In this work, the detection-rate of SO, CCS, OCS, and SO$_{2}$ are 100\% (50), 92\% (46), 96\% (48), and 42\% (21), respectively, 
which are summarized in Table~\ref{table:sm}. 
All these four species were simultaneously detected in 20 sources 
being good candidates to study the sulfur chemistry. 

\subsubsection{CH$_{3}$OH, CH$_{3}$OCH$_{3}$, CH$_{3}$OCHO, 
CH$_{3}$CHO, C$_{2}$H$_{5}$OH, and CH$_{3}$COCH$_{3}$}  \label{sec:ch3oh}  
CH$_{3}$OH, a key species in the formation of COMs,  
was exhaustively studied experimentally 
and found to be abundant and widespread in the ISM~\citep{Qin:22}. 
CH$_{3}$OCH$_{3}$, CH$_{3}$OCHO, 
CH$_{3}$CHO, and C$_{2}$H$_{5}$OH 
are related with CH$_{3}$OH in chemistry~\citep{Lefloch:17,Singh:19,Chen:23}. 
Their detection rates are 70\% (35), 64\% (32), 86\% (43), and 24\% (12), respectively, 
the detailed detection of which is given in Table~\ref{table:om}. 
Our sample would be appreciated for statistically analyzing the relations 
among CH$_{3}$OCH$_{3}$, CH$_{3}$OCHO, and CH$_{3}$CHO. 

CH$_{3}$COCH$_{3}$ is the simplest and smallest ketone, 
which was firstly detected in Sgr B2 by~\citet{Combes:87} and later confirmed by~\citet{Snyder:02}. 
The distribution of CH$_{3}$COCH$_{3}$ is clearly different from those of typical 
oxygen-bearing molecules (CH$_{3}$OH, CH$_{3}$OCHO) in Orion KL~\citep{Friedel:08,Peng:13}.
The differences between CH$_{3}$COCH$_{3}$ and other large oxygen-bearing
molecules make CH$_{3}$COCH$_{3}$ an excellent tool for testing different
chemical models of complex molecules, i.e., in the
gas phase by pure cosmic ray induced ion-molecule chemistry,
on the surface of grains, or both. 
The formation of CH$_{3}$COCH$_{3}$ in space is still unclear. 
CH$_{3}$COCH$_{3}$ was only detected in 4 of our target sources ($\sim$10\%), as shown in Table~\ref{table:om}. 
High resolution spectral line surveys of large sample are needed to tackle the issue systematically. 

\subsubsection{HCOOH and CH$_{3}$COOH} \label{sec:hcooh}
HCOOH, the simplest carboxylic acid, together with its descendant CH$_{3}$COOH~\citep{Pilling:11}, 
are possible precursors of prebiotic species~\citep{Fedoseev:15}. 
They have been extensively detected 
in the ISM~\citep{Schutte:99}, comets~\citep{Bennett:07} and meteorites~\citep{Bisschop:07}.   
The detection-rate of HCOOH is 72\% (36), 
while CH$_{3}$COOH was only detected in G010.47$+$00.02, 
G049.48$-$00.36, and G049.48$-$00.38, as summarized in Table~\ref{table:omh}. 

\subsubsection{HNCO and NH$_{2}$CHO}  \label{sec:hnco}
Both HNCO and NH$_{2}$CHO are potential prebiotic molecules 
that contain a peptide bond [$-$(H)N$-$C(O)$-$],  
which is the link of amino acids, and consequently form proteins. 
HNCO is the simplest molecule containing four of 
the six biogenic elements: carbon, hydrogen, oxygen, nitrogen, phosphorus, and sulfur 
that are present in all living beings, which has been identified in 43 out of our 50 target sources ($\sim$86\%). 
NH$_{2}$CHO, one of the simplest organic amide, has been suggested to be crucial 
in the synthesis of metabolic and genetic molecules, the chemical basis of life~\citep{Saladino:12}, 
and its detection rate is~50\% ($\sim$25 sources). 
The summary of detected HNCO and NH$_{2}$CHO is listed in Table~\ref{table:nm}. 
~\citet{Gorai:20} studied three nitrogen-bearing molecules (HNCO, NH$_{2}$CHO, and CH$_{3}$NCO) 
containing peptide-like bonds in G10.47$+$0.03 
and found that HNCO, NH$_{2}$CHO, and CH$_{3}$NCO are chemically linked with each other. The chemical links between HNCO and NH$_{2}$CHO were confirmed by~\citet{Taniguchi:23}.Therefore, our sample can be used to systematically study the link between HNCO and NH$_{2}$CHO, which would shield a light on the evolution of amino acids in the ISM. 

\subsubsection{HC$_{3}$N and HC$_{5}$N} \label{sec:hc3n} 
HC$_{3}$N, and HC$_{5}$N are two of carbon-chain species 
that are prone to be depleted onto dust grains 
when the gas is cold, and destroyed by UV radiations~\citep{Sakai:13}, 
the detected of which in our sample is presented in Table~\ref{table:nmh}. 
HC$_{3}$N and HC$_{5}$N were first detected in Sgr B2~\citep{Turner:71,Broten:76},  
and have been found to be widespread in the Milky Way ISM~\citep{Crovisier:04,Fontani:17,Taniguchi:19,Wang:22}. 
Their detection rates in our sample are 100\% (50) and 58\% (29), respectively.  

\subsubsection{CN, CH$_{3}$CN, C$_{2}$H$_{3}$CN, and C$_{2}$H$_{5}$CN} \label{sec:cn}
CN, the first observed nitrogen-bearing species in the ISM~\citep{McKellar:40}, 
can be produced from HCN via photodissociation and 
from reactions involving C$^{+}$ and C~\citep{Boger:05}.
CN can be thus taken as a probe of dense gas exposed to strong ultraviolet (UV) 
radiation in photon-dominated regions (PDRs).
Formation of CN at high column densities can be driven 
by enhanced cosmic ray ionization rates~\citep{Boger:05} 
or by X-rays near an active galactic nucleus (AGN) as shown by models of 
X-ray-dominated regions~\citep{Meijerink:07,Wolfire:22}. 
CN was detected in all sources of our sample, 
which combined with known PDRs deserves further study of dense gas properties. 

CH$_{3}$CN, C$_{2}$H$_{3}$CN, and C$_{2}$H$_{5}$CN 
were found to be related with CN~\citep{Zeng:18,Ribeiro:20}. 
CH$_{3}$CN is present in all sources except for G168.06$+$00.82, 
while C$_{2}$H$_{3}$CN and C$_{2}$H$_{5}$CN were observed 
in 16 ($\sim$32\%) and 20 ($\sim$40\%) sources of our sample, respectively. 
The relation between C$_{2}$H$_{3}$CN and C$_{2}$H$_{5}$CN 
can be studied with the target sources hosting them. 
We summarized the detected CN, CH$_{3}$CN, C$_{2}$H$_{3}$CN, 
and C$_{2}$H$_{5}$CN in Table~\ref{table:nmc}.  

\subsection{The detected ion HOCO$^{+}$} \label{sec:hoco}
\begin{figure*}
 \centering
  \includegraphics[width=0.65\textwidth]{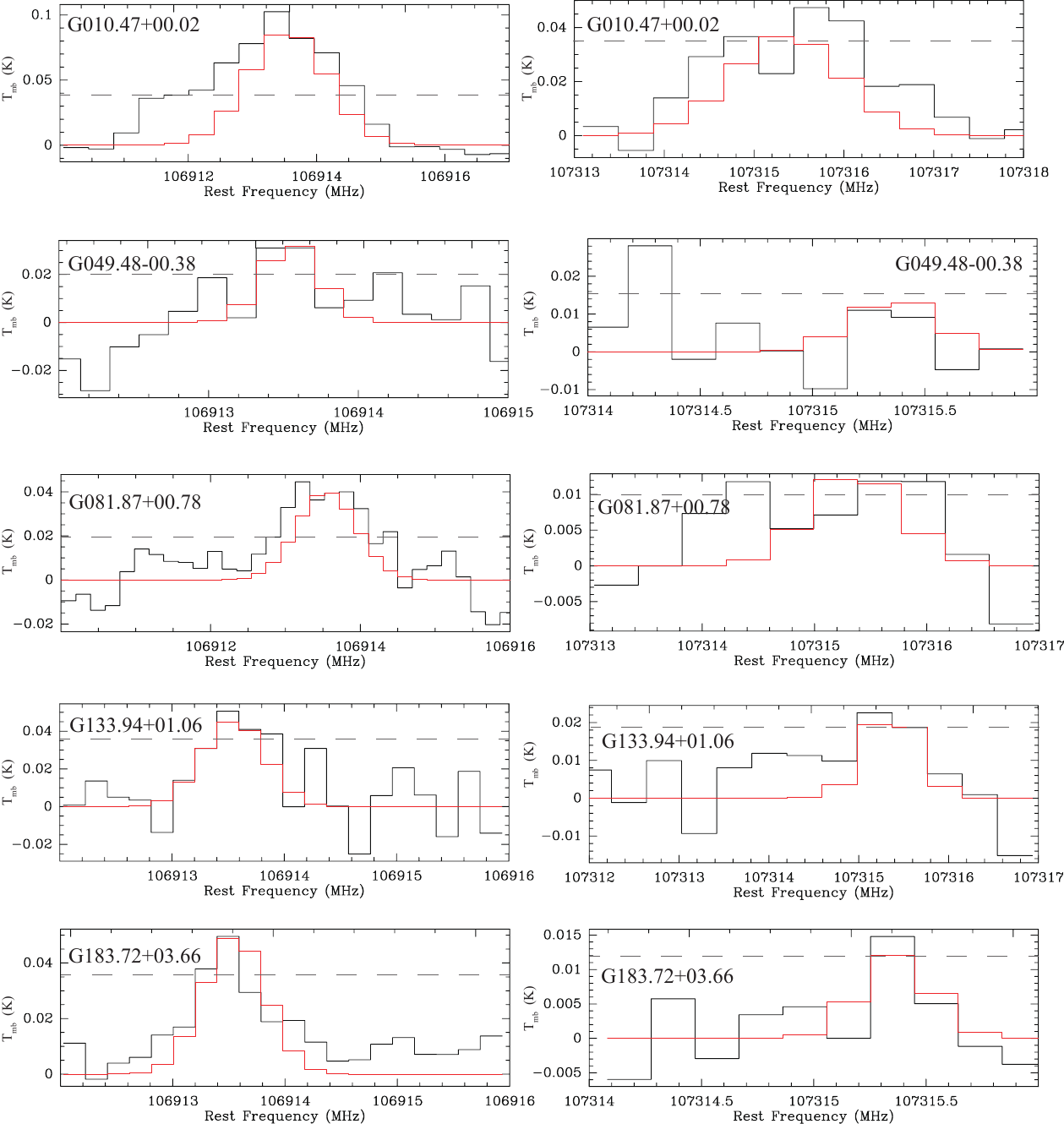}
   \caption{The detected two transitions of HOCO$^{+}$ in 5 target sources. 
   Black lines, red lines, and dash lines are observed lines, modeling synthetic spectrum, and 3$\sigma$ level, respectively.}
\label{fig:hoco}           
\end{figure*}

CO$_{2}$, lack of a permanent dipole moment, is an important reservoir of carbon and oxygen, 
and one of the major constituents of the icy mantles of dust grains.
The protonated form of CO$_{2}$, HOCO$^{+}$, 
is an interesting alternative to track the gas phase CO$_{2}$ in the millimeter/sub-millimeter regime.
~\citet{Fontani:18} first represented the statistical study of protonated carbon dioxide in 11 massive star-forming regions, 
and found that the ion-radical reaction $HCO^{+} + OH \rightarrow HOCO^{+} + H$ plays a more important role than 
a direct protonation of CO$_{2}$ in the gas-phase formation route of HOCO$^{+}$. 
Particularly, two transitions of HOCO$^{+}$, 
4$_{0,4}$ $-$ 3$_{0,3}$ (85531.497\,MHz) and 5$_{0,5}$ $-$ 4$_{0,4}$ (106913.545\,MHz), 
have been observed in IRAS 16293 $-$ 2422 to study the chemistry of CO$_{2}$ ices according to~\citet{Majumdar:18}.        
In our work, the transition 5$_{0,5}$ $-$ 4$_{0,4}$ also has been detected in G010.47$+$00.02 and G081.87$+$00.78, 
while it is marginally detected in G049.48$-$00.38, G133.94$+$01.06, and G183.72$+$03.66, graphically shown in Figure~\ref{fig:hoco}.  
Additionally, the transition 5$_{1,4}$ $-$ 4$_{1,3}$ (107315.356\,MHz) is weaker than 5$_{0,5}$ $-$ 4$_{0,4}$, 
and has been marginally detected in the five sources (Figure~\ref{fig:hoco}). 
Our target sources, where HOCO$^{+}$ was detected, enlarge the sample size to study the formation of HOCO$^{+}$, 
and therefore might advance the understanding of its interstellar chemistry.     

\subsection{Lines in vibrational state and with high E$_{u}$} \label{sec:liv} 
Molecular complexity in the ISM is not only manifest in increasingly larger molecules, 
but also in excited vibrational states of abundant molecules. 
The vibrational transitions generally have high upper level energy 
(E$_{u}$) compared with the ground state transitions, 
which emit radiation from the centimeter to sub-millimeter wavelength, 
unaffected by the dust extinction, 
making them powerful tools to probe the energetic processes 
taking place in the innermost regions of heavily obscured galactic nuclei~\citep{Rico:21}. 
On the other hand, combined with ground state lines, 
vibrational lines can be used to investigate the physical dynamic parameters, 
such as the gas distribution, velocity structure~\citep{Qin:15}, 
and the astro-chemical evolution of the sources where they reside.     
Vibrationally excited lines HC$_{3}$N v$_{7}$=1 were observed in 24 of our target sources,  
while HC$_{3}$N v$_{4}$=1, v$_{5}$=1/v$_{7}$=3, v$_{6}$=1, 
v$_{6}$=$_{7}$=1, and v$_{7}$=2, C$_{2}$H$_{5}$CN v$_{12}$=1-A and v$_{20}$=1-A 
were only detected in G10.47$+$00.02. 
CH$_{3}$CN v$_{8}$=1 and SO$_{2}$ v$_{2}$=1 were respectively identified in 
3 (G10.47$+$00.02, G049.48$-$00.36, and G049.48$-$00.38) 
and 4 (G10.47$+$00.02, G043.16$+$00.01, G049.48$-$00.36, and G049.48$-$00.38) target sources, 
as summarized in Table~\ref{table:lvh}. 
Therefore, G10.47$+$00.02, G043.16$+$00.01, G049.48$-$00.36, and G049.48$-$00.38 deserve to be 
systematically studied their astro-chemical evolution and physical dynamics with high sensitivity maps. 

\section{Summary} \label{sec:sum}
We have performed high-sensitivity broadband spectroscopic observations toward 50 massive 
star-forming regions with IRAM 30-m millimeter telescope. 
Weeds was used to identify molecules in each star-forming region. 
The related parameters, such as peak temperature, integrated intensity, and line width 
of the identified molecular lines were obtained. 
Even though no new molecule is discovered, the results provide a valuable dataset for further studies on the astro-chemical evolution of molecules in massive star-forming cores.
Our main results are as follows:
\begin{enumerate}
 \item There are 27 identified species, of which 16 are complex organic molecules. 
 \item Some chemically related molecules with high detection rate ($>$~60\%), such as CH$_{3}$OCH$_{3}$ ($\sim$64\%), CH$_{3}$OCHO ($\sim$64\%), and  CH$_{3}$CHO ($\sim$84\%), deserve to be statistically investigated their correlations and possible evolution pathways. 
 \item The line widths of the chemically related molecules exhibit minimal fluctuations, suggesting they likely originate from similar gases in star-forming regions.
\item The transition of HOCO$^{+}$, 5$_{0,5}$ $-$ 4$_{0,4}$ (106913.545\,MHz), 
is detected in G010.47$+$00.02 and G081.87$+$00.78, 
while it is marginally detected in G049.48$-$00.38, G133.94$+$01.06, and G183.72$+$03.66. 
\end{enumerate}  

\section*{Supplementary data}
The completed tables~\ref{table:paraweeds} and~\ref{table:moled} are available in the supplementary material.

\begin{ack}
This work is supported by the National Natural Science Foundation of China (grant No. 12103010 and U1931104), Project funded by China Postdoctoral Science Foundation (Grant No. 2022M710016), Fundamental Research Funds for the Central Universities (Grant No. 2020CDJXZ002), Chongqing Municipal Natural Science Foundation General Program (Grant No. cstc2021jcyj-msxmX0867), and Chongqing Talents: Exceptional Young Talents Project (Grant No. cstc2021ycjh-bgzxm0027). This study is based on observations carried out under project numbers 012-16 and 023-17 with the IRAM 30-m telescope. IRAM 30-m is supported by INSU/CNRS (France), MPG (Germany) and IGN (Spain).  
\end{ack}
\section*{Data Availability}
The original data observed with IRAM 30-m can be accessed by IRAM archive system
\footnote{https://www.iram-institute.org/EN/content-page-386-7-386-0-0-0.html}. 
If anyone is interested in the  reduced data presented in this paper, please contact Junzhi Wang at junzhiwang@gxu.edu.cn. 



\tiny
\begin{longtable}{*{6}{l}}  
\caption{Detected molecules and radio recombination line in sources.} \label{table:mole}\\    
\hline
\hline
Source & $<$ 6 Atoms & $\geqq$ 6 Atoms & Isotopic molecules & Radio recombination line & N$^{\ast}$ \\ 
\textbf{(1)} & \textbf{(2)} & \textbf{(3)} & \textbf{(4)} & \textbf{(5)} & \textbf{(6)} \\
\hline
\endfirsthead
\hline
\hline
Source & $<$ 6 Atoms & $\geqq$ 6 Atoms & Isotopic molecules & Radio recombination line & N$^{\ast}$ \\ 
\textbf{(1)} & \textbf{(2)} & \textbf{(3)} & \textbf{(4)} & \textbf{(5)} & \textbf{(6)} \\
\hline
\endhead
\endfoot
\multicolumn{6}{l}{{\bf Note.} Values of N$^{\ast}$ do not include radio recombination lines.}\\
\endlastfoot
{} & SO, SiS, CCS, & NH$_{2}$CHO, CH$_{3}$OH, HC$_{5}$N, & $^{34}$SO, OC$^{34}$S, & {} & {}\\  
{} & OCS, SO$_{2}$, & CH$_{3}$CN, C$_{2}$H$_{3}$CN,  & CH$_{3}$C$^{15}$N, & {} & {}\\ 
G000.67$-$00.03 & HNCO, HC$_{3}$N, & CH$_{3}$OCHO, C$_{2}$H$_{5}$CN, & $^{13}$CH$_{3}$CN, & H (39) $\alpha$, & 26 \\ 
{} & t$-$HCOOH & C$_{2}$H$_{5}$OH, CH$_{3}$CHO, & C$^{18}$O, $^{13}$CO, & H (48) $\beta$ & {}\\  
{} & {} & CH$_{3}$OCH$_{3}$ & CH$_{3}$$^{13}$CN,  C$^{17}$O & {} & {}\\ \hline
{} & SO, CN, CCS, & CH$_{3}$OH, CH$_{3}$CN, HC$_{5}$N, & $^{13}$CN,  C$^{18}$O, & H (39) $\alpha$, & {}\\
G005.88$-$00.39 & OCS, SO$_{2}$, & C$_{2}$H$_{3}$CN, CH$_{3}$OCHO, &  C$^{15}$N, $^{13}$CO,  & H (48) $\beta$, & 19 \\ 
{} & HNCO, HC$_{3}$N & CH$_{3}$C$_{3}$N, CH$_{3}$OCH$_{3}$, & C$^{17}$O & H (55) $\gamma$ & {} \\
\hline
\end{longtable}

\tiny
\setlength{\tabcolsep}{1.5pt}
\begin{longtable}{lcccccccc}  
\caption{Detected lines and their parameters in each sources} \label{table:moled}\\    
\hline
\hline 
Species & Transitions & Rest Freq. & E$_{u}$ & $\mu$$^{2}$S & T$_{mb}$ & 
V$_{\rm LSR}$ & $\Delta$V & $\int$ T$_{mb}$ dv \\
{} & {} & (MHz) & (K) & (D$^{2}$) & (mK) & (km~s$^{-1}$) & (km~s$^{-1}$) & (mK km~s$^{-1}$) \\
\textbf{(1)} & \textbf{(2)} & \textbf{(3)} & \textbf{(4)} & \textbf{(5)} & \textbf{(6)} & \textbf{(7)} & \textbf{(8)} & \textbf{(9)}\\
\hline
\endfirsthead
\hline
\hline
Species & Transitions & Rest Freq. & E$_{u}$ & $\mu$$^{2}$S & T$_{mb}$ & 
V$_{\rm LSR}$ & $\Delta$V & $\int$ T$_{mb}$ dv \\
{} & {} & (MHz) & (K) & (D$^{2}$) & (mK) & (km~s$^{-1}$) & (km~s$^{-1}$) & (mK km~s$^{-1}$) \\
\textbf{(1)} & \textbf{(2)} & \textbf{(3)} & \textbf{(4)} & \textbf{(5)} & \textbf{(6)} & \textbf{(7)} & \textbf{(8)} & \textbf{(9)}\\
\hline
\endhead
\endfoot
\hline
\multicolumn{9}{l}{{\bf Note.} This is only a subset of the whole table, which is available in the supplementary material.} \\  
\multicolumn{9}{l}{~~~~~~~~~ (1): Molecule name; (2): Transition quantum numbers; (3): Rest frequency; 
(4): Upper state energy level (K); } \\ 
\multicolumn{9}{l}{~~~~~~~~~ (5): Dipole-weighted transition dipole matrix elements; 
(6): Peak intensity; (7): centroid velocity;
(8): FWHM,} \\
\multicolumn{9}{l}{~~~~~~~~~ (9): Integrated intensity. "$-$" represents no values.}\\
\endlastfoot
{} & {} & {} & {} & {} & {} & {} & {} & {} \\
\multicolumn{9}{c}{G000.67$-$00.03}\\
{} & {} & {} & {} & {} & {} & {} & {} & {} \\
NH$_{2}$CHO & 5(2, 4) $-$ 4(2, 3) & 105972.665(37e-3) & 27.2 & 54.915 & 611 & 62.5(0.1) & 13.2 (0.3) & 8594(140)\\
NH$_{2}$CHO & 5(4, 1) $-$ 4(4, 0) & 106107.870(88e-3) & 63.0 & 23.537 & 239& 62.5(0.3) & 9.3(0.7) & 2360(138)\\
NH$_{2}$CHO & 5(3, 3) $-$ 4(3, 2) & 106134.468(55e-3) & 42.1 & 41.845 & 391 & 62.5(0.5) & 12.8(0.5) & 5319(171)\\
NH$_{2}$CHO & 5(3, 2) $-$ 4(3, 1) & 106141.442(55e-3) & 42.1 & 41.84 & 375 & 62.5(0.5) & 12.0(0.5) & 4791(160)\\
NH$_{2}$CHO & 5(2, 3) $-$ 4(2, 2) & 106541.773(37e-3) & 27.2 & 54.915 & 598 & 62.5(0.1) & 12.8(0.3) & 8163(172)\\
\end{longtable}

\begin{table*}
\tbl{Detected H (39) $\alpha$ and He (39) $\alpha$.}{
\centering
\begin{tabular}{lccccc}    
\hline
\hline
Sources & RRL & Rest Freq. & V$_{\rm LSR}$ & $\Delta$V & $\int$ T$_{mb}$ dv \\
{} & {} & (MHz) & (km~s$^{-1}$) & (km~s$^{-1}$) & (mK km~s$^{-1}$) \\
\hline
G000.67$-$00.03 & H (39) $\alpha$ & 106737.357(0) & 55.6(0.6) & 40.8(1.4) & 10701 (294)\\
G005.88$-$00.39 & H (39) $\alpha$ &106737.357(0) & 6.9(0.2) & 59.1(0.6) & 67990(543)\\ \hline
G009.62$+$00.19 & H (39) $\alpha$ &106737.357(0) & 7.9(0.7) & 19.6(1.8) & 1342(98) \\ 
{} & He (39) $\alpha$ & 106780.852(0) & 8.2(1.0) & 9.2(1.8) & 319(63)\\  \hline 
G010.62$-$00.38 & H (39) $\alpha$ &106737.357(0) & 2.3(0.1) & 45.0(2.5) & 12209(1590) \\ 
{} & He (39) $\alpha$ & 106780.852(0) & 2.3(0.1) & 45.0(2.5) & 12209(1590)\\
\hline 
\end{tabular}}\label{table:RRLs}
\end{table*}

\setlength{\tabcolsep}{3pt} 
\begin{longtable}{lccc|lccc}  
\caption{Observational parameters of the remaining unidentified lines.}\label{table:rul}\\    
\hline
\hline
Sources & $\nu_{\rm obs}$ & $\Delta$V & $\int$ T$_{\rm mb}$ dv & Sources & $\nu_{\rm obs}$ & $\Delta$V & $\int$ T$_{\rm mb}$ dv \\
{} & (MHz) & (km~s$^{-1}$) & (mK km~s$^{-1}$) & {} & (MHz) & (km~s$^{-1}$) & (mK km~s$^{-1}$) \\
\hline
\endfirsthead
\hline
Sources & $\nu_{\rm obs}$ & $\Delta$V & $\int$ T$_{\rm mb}$ dv & Sources & $\nu_{\rm obs}$ & $\Delta$V & $\int$ T$_{\rm mb}$ dv \\
{} & (MHz) & (km~s$^{-1}$) & (mK km~s$^{-1}$) & {} & (MHz) & (km~s$^{-1}$) & (mK km~s$^{-1}$) \\
\hline
\endhead
{} & {} & {} & {} & G049.48$-$0.36 & 111503.390 & 5.9(1.3) & 418(63)\\    
G000.67$-$00.03 & 107606.336 & 9.6(2.3) & 595(128) & {} & 111518.493 & 8.9(1.4) & 503(68)\\                
{} & 107623.459 & 13.3(2.0) & 758(126) & {} & 112579.794 & 1.9(1.0) & 109(41)\\                
{} & 110607.218 & 4.8(1.5) & 564(128)  & {} & 112581.764 & 4.2(0.8) & 362(52)\\ \cline{5-8}               
{} & 111322.089 & 13.4(1.9) & 1767(195) & G049.48$-$00.38 & 105869.777 & 9.6(3.1) & 1009(107)\\ 
{} & 111347.774 & 6.0(1.7) & 491(125) & {} & 105882.363 & 7.1(2.1) & 205(52)\\                
{} & 111384.760 & 9.3(1.6) & 952(150) & {} & 105894.692 & 6.1(0.9) & 460(53)\\                
{} & 111600.856 & 1.2(0.8) & 132(59) & {} & 106333.887 & 7.2(0.5) & 355(19)\\
\hline  
\end{longtable}

\begin{table*}
	\tbl{The detection rate of the related molecules}{
		\centering
		\begin{tabular}{l|l}     
			\hline
			\hline
			Section & The related molecules with their detection rates \\  
			\hline
			Sect.~\ref{sec:so} & SO (100\%), CCS (92\%), OCS (96\%), SO$_{2}$ (42\%)\\  
			Sect.~\ref{sec:ch3oh} & CH$_{3}$OH (100\%), CH$_{3}$CHO (86\%), CH$_{3}$OCHO (64\%), CH$_{3}$OCH$_{3}$ (70\%), C$_{2}$H$_{5}$OH (24\%), CH$_{3}$COCH$_{3}$ (10\%)  \\
			Sect.~\ref{sec:hcooh} & HCOOH (72\%), CH$_{3}$COOH (6\%) \\
			Sect.~\ref{sec:hnco} & HNCO (86\%), NH$_{2}$CHO (50\%) \\
			Sect.~\ref{sec:hc3n} & HC$_{3}$N (100\%), HC$_{5}$N(58\%) \\
			Sect.~\ref{sec:cn} & CN (100\%), CH$_{3}$CN (98\%), C$_{2}$H$_{3}$CN (32\%), C$_{2}$H$_{5}$CN (40\%) \\
			\hline
    	\end{tabular}}\label{table:dr}
\end{table*}

\begin{table*}
\tbl{Detected SO, CCS, OCS, and SO$_{2}$ summary.}{
\centering
\begin{tabular}{lcccc|lcccc}    
\hline
\hline
Source & SO & CCS & OCS & SO$_{2}$ & Source & SO & CCS & OCS & SO$_{2}$ \\  
\hline
G000.67$-$00.03 & $\surd$ & $\surd$ & $\surd$ & $\surd$  &   G035.20$-$01.73 & $\surd$ & $\surd$ & $\surd$ & $\times$ \\  
G005.88$-$00.39 & $\surd$ & $\surd$ & $\surd$ & $\surd$  &   G037.43$+$01.51 & $\surd$ & $\surd$ & $\surd$ & $\surd$ \\   
G009.62$+$00.19 & $\surd$ & $\surd$ & $\surd$ & $\times$ &   G043.16$+$00.01 & $\surd$ & $\surd$ & $\surd$ & $\surd$ \\   
G010.47$+$00.02 & $\surd$ & $\surd$ & $\surd$ & $\surd$  &   G043.79$-$00.12 & $\surd$ & $\surd$ & $\surd$ & $\surd$ \\   
G010.62$-$00.38 & $\surd$ & $\surd$ & $\surd$ & $\surd$  &   G049.48$-$00.36 & $\surd$ & $\surd$ & $\surd$ & $\surd$ \\   
G011.49$-$01.48 & $\surd$ & $\times$ & $\times$ & $\times$ & G049.48$-$00.38 & $\surd$ & $\surd$ & $\surd$ & $\surd$ \\   
G011.91$-$00.61 & $\surd$ & $\surd$ & $\surd$ & $\times$  &  G059.78$+$00.06 & $\surd$ & $\surd$ & $\surd$ & $\times$ \\
\hline
\end{tabular}}\label{table:sm}
\begin{tabnote}
{\bf Note.} $\surd$ and $\times$ represent detection and non-detection for molecules, respectively.
\end{tabnote}
\end{table*}

\begin{table*}
\tbl{Detected CH$_{3}$OH, CH$_{3}$CHO, CH$_{3}$OCHO, 
CH$_{3}$OCH$_{3}$, C$_{2}$H$_{5}$OH, and CH$_{3}$COCH$_{3}$ summary.}{
\centering
\begin{tabular}{lcccccc}     
\hline
\hline
Source & CH$_{3}$OH & CH$_{3}$CHO & CH$_{3}$OCHO & CH$_{3}$OCH$_{3}$ & 
 C$_{2}$H$_{5}$OH & CH$_{3}$COCH$_{3}$ \\
\hline
G000.67$-$00.03 & $\surd$ & $\surd$ & $\surd$ & $\surd$ & $\surd$ & $\times$ \\
G005.88$-$00.39 & $\surd$ & $\times$ & $\surd$ & $\surd$ & $\times$ & $\times$ \\
G009.62$+$00.19 & $\surd$ & $\surd$ & $\surd$ & $\surd$ & $\surd$ & $\times$ \\
G010.47$+$00.02 & $\surd$ & $\surd$ & $\surd$ & $\surd$ & $\surd$ & $\surd$ \\
G010.62$-$00.38 & $\surd$ & $\surd$ & $\surd$ & $\surd$ & $\surd$ & $\times$ \\
G011.49$-$01.48 & $\surd$ & $\times$ & $\times$ & $\surd$ & $\times$ & $\times$ \\
G011.91$-$00.61 & $\surd$ & $\surd$ & $\surd$ & $\times$ & $\times$ & $\times$ \\
\hline
\end{tabular}}\label{table:om}
\begin{tabnote}
{\bf Note.} $\surd$ and $\times$ represent detection and non-detection for molecules, respectively.
\end{tabnote}
\end{table*}

\begin{table*}
\tbl{Detected HCOOH and CH$_{3}$COOH summary.}{
\centering
\begin{tabular}{lcc|lcc}     
\hline
\hline
Sources & HCOOH & CH$_{3}$COOH & Sources & HCOOH & CH$_{3}$COOH \\
\hline
G000.67$-$00.03 & $\surd$ & $\times$ & G037.43$+$01.51 & $\surd$ & $\times$ \\  
G009.62$+$00.19 & $\surd$ & $\times$ & G043.16$+$00.01 & $\surd$ & $\times$ \\  
G010.47$+$00.02 & $\surd$ & $\surd$  & G043.79$-$00.12 & $\surd$ & $\times$ \\  
G010.62$-$00.38 & $\surd$ & $\times$ & G049.48$-$00.36 & $\surd$ & $\surd$ \\   
G011.91$-$00.61 & $\surd$ & $\times$ & G049.48$-$00.38 & $\surd$ & $\surd$ \\   
G012.80$-$00.20 & $\surd$ & $\times$ & G059.78$+$00.06 & $\surd$ & $\times$ \\  
G012.88$+$00.48 & $\surd$ & $\times$ & G069.54$-$00.97 & $\surd$ & $\times$ \\  
G014.33$-$00.64 & $\surd$ & $\times$ & G075.76$+$00.33 & $\surd$ & $\times$ \\  
G023.44$-$00.18 & $\surd$ & $\times$ & G081.75$+$00.59 & $\surd$ & $\times$ \\  
G027.36$-$00.16 & $\surd$ & $\times$ & G081.87$+$00.78 & $\surd$ & $\times$ \\  
G028.86$+$00.06 & $\surd$ & $\times$ & G092.67$+$03.07 & $\surd$ & $\times$ \\  
G029.95$-$00.01 & $\surd$ & $\times$ & G109.87$+$02.11 & $\surd$ & $\times$ \\  
G031.28$+$00.06 & $\surd$ & $\times$ & G111.54$+$00.77 & $\surd$ & $\times$ \\  
G032.04$+$00.05 & $\surd$ & $\times$ & G121.29$+$00.65 & $\surd$ & $\times$ \\  
G034.39$+$00.22 & $\surd$ & $\times$ & G123.06$-$06.30 & $\surd$ & $\times$ \\  
G035.02$+$00.34 & $\surd$ & $\times$ & G133.94$+$01.06 & $\surd$ & $\times$ \\  
G035.19$-$00.74 & $\surd$ & $\times$ & G188.94$+$00.88 & $\surd$ & $\times$ \\  
G035.20$-$01.73 & $\surd$ & $\times$ & G192.60$-$00.04 & $\surd$ & $\times$ \\ 
\hline
\end{tabular}}\label{table:omh}
\begin{tabnote}
{\bf Note.} $\surd$ and $\times$ represent detection and non-detection for molecules, respectively.
\end{tabnote}
\end{table*}

\begin{table*}
\tbl{Detected HNCO and NH$_{2}$CHO summary.}{
\centering
\begin{tabular}{lcc|lcc}     
\hline
\hline
Sources & HNCO & NH$_{2}$CHO & Sources & HNCO & NH$_{2}$CHO \\
\hline
G000.67$-$00.03 & $\surd$ & $\surd$  & G035.02$+$00.34 & $\surd$ & $\times$ \\
G005.88$-$00.39 & $\surd$ & $\times$ & G035.19$-$00.74 & $\surd$ & $\surd$ \\ 
G009.62$+$00.19 & $\surd$ & $\surd$  & G035.20$-$01.73 & $\surd$ & $\times$ \\
G010.47$+$00.02 & $\surd$ & $\surd$  & G037.43$+$01.51 & $\surd$ & $\times$ \\
G010.62$-$00.38 & $\times$ & $\surd$ & G043.16$+$00.01 & $\surd$ & $\times$ \\
G011.91$-$00.61 & $\surd$ & $\surd$  & G043.79$-$00.12 & $\surd$ & $\surd$ \\ 
G012.80$-$00.20 & $\surd$ & $\surd$  & G049.48$-$00.36 & $\surd$ & $\surd$ \\ 
G012.88$+$00.48 & $\surd$ & $\surd$  & G049.48$-$00.38 & $\surd$ & $\surd$ \\ 
G012.90$-$00.24 & $\surd$ & $\times$ & G059.78$+$00.06 & $\surd$ & $\times$ \\
G012.90$-$00.26 & $\surd$ & $\surd$  & G069.54$-$00.97 & $\surd$ & $\times$ \\
G014.33$-$00.64 & $\surd$ & $\surd$  & G078.12$+$03.63 & $\surd$ & $\surd$ \\ 
G015.03$-$00.67 & $\surd$ & $\times$ & G081.75$+$00.59 & $\surd$ & $\times$ \\
G016.58$-$00.05 & $\surd$ & $\times$ & G081.87$+$00.78 & $\surd$ & $\surd$ \\ 
G023.00$-$00.41 & $\surd$ & $\surd$  & G092.67$+$03.07 & $\surd$ & $\times$ \\
G023.44$-$00.18 & $\surd$ & $\surd$  & G109.87$+$02.11 & $\surd$ & $\times$ \\
G027.36$-$00.16 & $\surd$ & $\surd$  & G111.54$+$00.77 & $\surd$ & $\surd$ \\ 
G028.86$+$00.06 & $\surd$ & $\surd$  & G121.29$+$00.65 & $\surd$ & $\times$ \\
G029.95$-$00.01 & $\surd$ & $\surd$  & G123.06$-$06.30 & $\surd$ & $\times$ \\
G031.28$+$00.06 & $\surd$ & $\surd$  & G133.94$+$01.06 & $\surd$ & $\times$ \\
G031.58$+$00.07 & $\surd$ & $\times$ & G192.60$-$00.04 & $\surd$ & $\surd$ \\ 
G032.04$+$00.05 & $\surd$ & $\surd$  & G209.00$-$19.38 & $\surd$ & $\times$ \\
G034.39$+$00.22 & $\surd$ & $\surd$  & G232.62$+$00.99 & $\surd$ & $\times$ \\
\hline
\end{tabular}}\label{table:nm}
\begin{tabnote}
{\bf Note.} $\surd$ and $\times$ represent detection and non-detection for molecules, respectively.
\end{tabnote}
\end{table*} 

\begin{table*}
\tbl{Detected HC$_{3}$N and HC$_{5}$N summary.}{
\centering
\begin{tabular}{lcc|lcc}     
\hline
\hline
Sources & HC$_{3}$N & HC$_{5}$N & Sources & HC$_{3}$N & HC$_{5}$N \\
\hline
G000.67$-$00.03 & $\surd$ & $\surd$  & G035.20$-$01.73 & $\surd$ & $\times$ \\
G005.88$-$00.39 & $\surd$ & $\surd$  & G037.43$+$01.51 & $\surd$ & $\surd$ \\ 
G009.62$+$00.19 & $\surd$ & $\surd$  & G043.16$+$00.01 & $\surd$ & $\times$ \\
G010.47$+$00.02 & $\surd$ & $\surd$  & G043.79$-$00.12 & $\surd$ & $\times$ \\
G010.62$-$00.38 & $\surd$ & $\surd$  & G049.48$-$00.36 & $\surd$ & $\surd$ \\ 
G011.49$-$01.48 & $\surd$ & $\times$ & G049.48$-$00.38 & $\surd$ & $\surd$ \\  
G011.91$-$00.61 & $\surd$ & $\surd$  & G059.78$+$00.06 & $\surd$ & $\surd$ \\
\hline
\end{tabular}}\label{table:nmh}
\begin{tabnote}
{\bf Note.} $\surd$ and $\times$ represent detection and non-detection for molecules, respectively.
\end{tabnote}
\end{table*}

\begin{table*}
\tbl{Detected CN, CH$_{3}$CN, C$_{2}$H$_{3}$CN and C$_{2}$H$_{5}$CN summary.}{
\centering
\begin{tabular}{lcccc|lcccc}     
\hline
\hline
Sources & CN & CH$_{3}$CN & C$_{2}$H$_{3}$CN & C$_{2}$H$_{5}$CN & 
Sources & CN & CH$_{3}$CN & C$_{2}$H$_{3}$CN & C$_{2}$H$_{5}$CN \\
\hline
G000.67$-$00.03 & $\surd$ & $\surd$ & $\surd$ & $\surd$  &  G035.20$-$01.73 & $\surd$ & $\surd$ & $\times$ & $\times$ \\ 
G005.88$-$00.39 & $\surd$ & $\surd$ & $\surd$ & $\times$  & G037.43$+$01.51 & $\surd$ & $\surd$ & $\times$ & $\times$ \\  
G009.62$+$00.19 & $\surd$ & $\surd$ & $\surd$ & $\surd$  &  G043.16$+$00.01 & $\surd$ & $\surd$ & $\times$ & $\surd$ \\  
G010.47$+$00.02 & $\surd$ & $\surd$ & $\surd$ & $\surd$  &  G043.79$-$00.12 & $\surd$ & $\surd$ & $\times$ & $\surd$ \\  
G010.62$-$00.38 & $\surd$ & $\surd$ & $\surd$ & $\surd$  &  G049.48$-$00.36 & $\surd$ & $\surd$ & $\surd$ & $\surd$ \\   
G011.49$-$01.48 & $\surd$ & $\surd$ & $\times$ & $\times$ & G049.48$-$00.38 & $\surd$ & $\surd$ & $\surd$ & $\surd$ \\    
G011.91$-$00.61 & $\surd$ & $\surd$ & $\surd$ & $\surd$  &  G059.78$+$00.06 & $\surd$ & $\surd$ & $\times$ & $\times$ \\
\hline
\end{tabular}}\label{table:nmc}
\begin{tabnote}
{\bf Note.} $\surd$ and $\times$ represent detection and non-detection for molecules, respectively.
\end{tabnote}
\end{table*}

\begin{table*}
\tbl{Detected HC$_{3}$N, SO$_{2}$, CH$_{3}$CN and C$_{2}$H$_{5}$CN in vibrational state.}{
\centering
\begin{tabular}{lcccccc}     
\hline
\hline
Sources & HC$_{3}$N & HC$_{3}$N other & SO$_{2}$ & CH$_{3}$CN & C$_{2}$H$_{5}$CN & C$_{2}$H$_{5}$CN \\
{} & v$_{7}$=1 & vibrational states & v$_{2}$=1 & v$_{8}$=1 & v$_{12}$=1-A & v$_{20}$=1-A \\
\hline
G000.67$-$00.03 & $\surd$ & $\times$ & $\times$ & $\times$ & $\times$ & $\times$ \\ 
G005.88$-$00.39 & $\surd$ & $\times$ & $\times$ & $\times$ & $\times$ & $\times$ \\ 
G009.62$+$00.19 & $\surd$ & $\times$ & $\times$ & $\times$ & $\times$ & $\times$ \\ 
G010.47$+$00.02 & $\surd$ & $\surd$ & $\surd$ & $\surd$ & $\surd$ & $\surd$ \\ 
G010.62$-$00.38 & $\surd$ & $\times$ & $\times$ & $\times$ & $\times$ & $\times$ \\ 
G011.91$-$00.61 & $\surd$ & $\times$ & $\times$ & $\times$ & $\times$ & $\times$ \\
\hline
\end{tabular}}\label{table:lvh}
\begin{tabnote}
{\bf Note.} $\surd$ and $\times$ represent detection and non-detection for molecules, respectively.
\end{tabnote}
\end{table*}

\end{document}